\documentclass[a4paper,UKenglish,cleveref, autoref]{lipics-v2019}
\nolinenumbers 
\usepackage{amsmath}
\usepackage{listings}
\usepackage{latexsym}
\usepackage{graphicx}
\usepackage{url}       % SB
\usepackage{algorithmic}
\usepackage{algorithm}
\usepackage{times}
\usepackage{color}
\usepackage{semantic}
\usepackage{stmaryrd}
\usepackage{amssymb}
\usepackage{textcomp}
\usepackage{tikz}

\makeatletter
\newcommand*\bigcdot{\mathpalette\bigcdot@{.5}}
\newcommand*\bigcdot@[2]{\mathbin{\vcenter{\hbox{\scalebox{#2}{$\m@th#1\bullet$}}}}}
\makeatother
\newcommand{\wand}{\mathrel{-\hspace{-.7ex}*}}
\newcommand{\triple}[3]{\{#1\}\,#2\,\{#3\}}
\newcommand{\denote}[1]{\llbracket #1 \rrbracket}
\newcommand{\defeq}{=_{\mathrm{def}}}

\newcommand{\listrep}{ \hspace{0.2em} {\vspace{-0.1ex}\shortmid\vspace{0.1ex}} \hspace{-0.35em} \leadsto}
\newcommand{\lseg}[1]{ \stackrel{#1}{{\vspace{-0.1ex}\shortmid\vspace{0.1ex}} \hspace{-0.35em} \leadsto \hspace{-0.45em} {\vspace{-0.1ex}\shortmid\vspace{0.1ex}}}  }
\newcommand{\ins}{\mathrm{ins}}
\newcommand{\fieldadd}[2]{{#1}.{#2}}
\title{Proof Pearl: Magic Wand as Frame}
\author{Qinxiang Cao}{Shanghai Jiao Tong University}{caoqinxaing@gmail.com}{}{}
\author{Shengyi Wang}{National University of Singapore}{}{}{}
\author{Aquinas Hobor}{National University of Singapore}{}{}{}
\author{Andrew W. Appel}{Princeton University}{}{}{}
\authorrunning{Q. Cao et al.}
\Copyright{Qinxiang Cao, Shengyi Wang, Aquinas Hobor and Andrew W. Appel}

\begin{document}

\maketitle

\begin{abstract}
	
Separation logic adds two connectives to assertion languages: separating conjunction $*$ (``star'') and its adjoint, 
separating implication $\wand$ (``magic wand'').
Comparatively, separating implication is less widely used.

This paper demonstrates that by using magic wand to express \emph{frame}s that relate 
\emph{mutable} local portions of data structures to global portions, we can exploit its power while proofs are still easily understandable. 
Many useful separation logic theorems about partial data structures can now be proved by simple automated tactics, which were usually proved by induction.
This magic-wand-as-frame technique is especially useful when formalizing the proofs by a high order logic.
We verify binary search tree insert in Coq as an example to demonstrate this proof technique.

\end{abstract}

\section{Introduction}
Separation logic \cite{reynolds02} is an extension of Hoare logic
that has been widely used in program verification.
The \emph{separating conjunction} $P * Q$ (``star'') in assertions represents the existence of two disjoint states, one that satisfies $P$ and one that satisfies $Q$.
Formally,
\[
\begin{array}{l}
m \vDash P*Q \ \ \defeq \ \
\text{there exist $m_1$ and $m_2$ s.t. } \\
\ \ \ \ \ \ \ \ \ \ \ \ \ \ \ \ \ \ \ \ \ \ \ \ \ \ \ \ \ \ \ \ \ \ \text{$m = m_1 \oplus m_2$, $m_1 \vDash P$ and $m_2\vDash Q$.}
\end{array}
\]
Here, $m_1 \oplus m_2$ represents the disjoint union of two pieces of state/memory.
The $*$ concisely expresses
address (anti)\-al\-ias\-ing.
For example, if ``$p \mapsto v$'' is the assertion that data $v$ is stored at address $p$, then $~p \mapsto v * q \mapsto u~$
says $v$ is stored at address $p$, $u$ is stored at address $q$, and $p \neq q$.
Separation logic enables one to verify a Hoare triple locally but use it globally, using the \emph{frame rule:}
\[
\inference[\textsc{frame}]{\triple{P}{c}{Q} & \mathrm{FV}(F) \cap \mathrm{ModV}(c) = \emptyset}{\triple{P * F}{c}{Q*F}}
\]
Star has a right adjoint $P \wand Q$ \emph{separating implication}, a.k.a. ``magic wand'':
\begin{align*}
&\begin{array}{l}
m \vDash P \wand Q \ \ \defeq \ \ 
\text{for any $m_1$ and $m_2$,} \\
\ \ \ \ \ \ \ \ \ \ \ \ \ \ \ \ \ \ \ \ \ \ \ \ \ \ \ \text{if $m \oplus m_1 = m_2$ and $m_1 \vDash P$ then $m_2\vDash Q$.} 
\end{array}
\end{align*}
And the following rules are called adjoint rules:
\begin{align*}
&\inference[\textsc{wand-adjoint1}]{P \vdash Q \wand R}{P * Q \vdash R} ~~~~~~~~ \inference[\textsc{wand-adjoint2}]{P * Q \vdash R}{P \vdash Q \wand R}
\end{align*}

Magic wands are famously difficult to control~\cite{rowling97}.
In the early days of separation logic, magic wand was used to generate weakest preconditions and verification conditions for automated program verification. However, those verification conditions are not human readable or understandable, and decision procedures for entailment checking with magic wand are quite complex. %TODO: CITE

In most works of interactive program verification, magic wand is not a necessary component.
Authors tend to use forward verification instead of backward verification since it is easier to understand a program correctness proof that goes in the same direction
as program execution. ``Forward'' Hoare logic rules do not generate magic wand expressions; therefore, most authors find that the expressive power of star is already strong enough. For example, we need to define separation logic predicates for different data structures (like records, arrays, linked list and binary trees) in order to verify related programs. Berdine \emph{et al.} \cite{berdine2004decidable} and
Chargu\'eraud \cite{chargueraud2016higher} show that these predicates can be defined with separating conjunction only.

\begin{figure*}[h]
\hspace{.2in}
\begin{minipage}{2.8in}
\begin{lstlisting}[language=C,backgroundcolor = \color{white}]
$~$
struct tree {int key; void *value;
             struct tree *left, *right;};

typedef struct tree **treebox;

void insert (treebox p, int x, void *v) {
  struct tree *q;
  while (1) {
  q = *p;
  if (q==NULL) {
    q = (struct tree *) surely_malloc (sizeof *p);
    q$\to$key=x; q$\to$value=v;
    q$\to$left=NULL; q$\to$right=NULL;
    *p=q;
    return;
  } else {
    int y = q$\to$key;
    if (x<y)
      p= &q$\to$left;
    else if (y<x)
      p= &q$\to$right;
    else {
      q$\to$value=v;
      return;
} } } }
$~$
\end{lstlisting}
\vspace{0.1in}
\caption{Binary Search Tree insertion}
\label{insert-program}
 %Dashed line shows $\mathrm{Fr}_{t, p}^{t_0, p_0}$}
\end{minipage}
\begin{minipage}{2.5in}
\hspace{.5in}
\begin{tabular}{c}
\includegraphics[scale=.8]{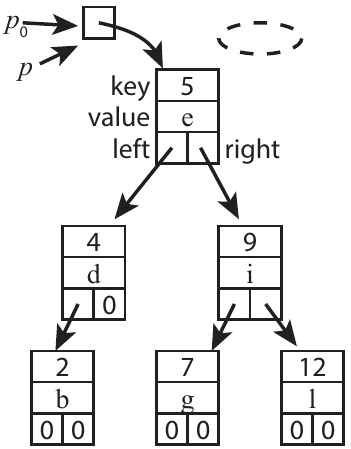}\\[2ex]
\includegraphics[scale=.8]{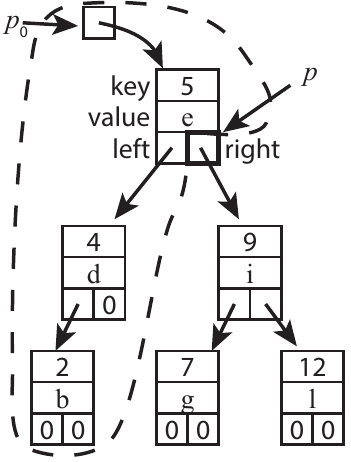}\\[2ex]
\includegraphics[scale=.8]{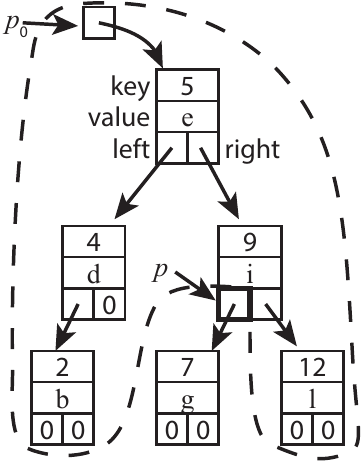}
\end{tabular}
\vspace{-0.15in}
\caption{Execution of $\mathsf{insert}(p_0,8,\mbox{\lstinline{"h"}})$. }
\label{fig-exec}
\end{minipage}
\vspace{.2in}
\end{figure*}

In this paper, we propose a proof technique: \emph{magic wand as frame}. Specifically, we propose to use magic wand together with quantifiers to define separation logic predicates for partial data structures.

The main content of this paper is a proof pearl. We use our magic-wand-as-frame technique to verify the C program in Fig.~\ref{insert-program}, insertion into a binary search tree (BST).
The program uses a while loop to walk down from the root to the location to insert the new element.

A pointer to a tree has type \lstinline{struct tree *},
but we also need the type pointer-to-pointer-to-tree, which we call
\lstinline{treebox}.  The \lstinline{insert} function does not return a new
tree, it modifies the old tree, perhaps replacing it entirely (if the
old tree were the empty tree).

Consider running \lstinline{insert($p_0$,8,``h")},
where $p_0$ points to a treebox containing the root of a tree as shown in
Fig. \ref{fig-exec}.
After one iteration of the loop or two iterations,
variable $\mathsf{p}$ contains
address $p$, which is a treebox containing a pointer
to a subtree. This subtree $t$ and a partial tree $P$ (shown within the dashed line) form the original BST.

Naturally, we can verify such a program using a loop invariant with the following form:
$$~~~~P \textit{ is stored in memory} ~~~*~~~ t\textit{ is stored in memory} $$

To describe partial trees,
most authors \cite{berdine2004decidable,chargueraud2016higher}
would have you introduce a
new inductive description of
\emph{tree with exactly one hole}---in addition
to the inductive description of ordinary trees---and
define a corresponding recursive separation-logic predicate
``partial tree $P \textit{ is stored in memory}$'',
in addition to the recursive predicate for ordinary trees.
(Similarly, ``list segment'' is inductively defined as a list with one hole.)

That's a lot of duplication.
We propose a different approach in this paper: using magic wand and universal quantifier to express ``$P \textit{ is stored in memory}$''.
Specifically, it is defined as: $$\forall t^{*},  ~~~ ( ~~~ t^{*}\textit{ is stored in memory}  ~~~~ \wand ~~~~ P(t^{*})\textit{ is stored in memory} ~~~ ) $$
in which $P(t^{*})$ represents the result of filling the hole in $P$ with $t^{*}$.

It has two benefits. (1) Important properties of the partial-tree predicate can now be proved by \textsc{wand-adjoint} directly. In our Coq formalization, most of these proofs have only a simple line of tacitcs. In comparions, similar properties were usually proved by induction which are hard to automate.
(2) We do not even need to define ``partial trees'' as a new inductive type in Coq. Instead, we treat partial trees as functions from trees to trees. As a result, we get equations like $P_1 (P_2 (t)) = (P_1 \circ P_2) (t)$ for free.

We organize the rest of this paper as follows: 
\begin{itemize}
\item[\S\ref{sec:paper_proof}] We verify this C implementation of BST insert using magic-wand-as-frame.

\item[\S\ref{sec:Coq_proof}] We formalize this correctness proof in Coq using \emph{Verifiable C} \cite{appel14:plcc}.

\item[\S\ref{sec:other_proof}] We show that magic-wand-as-frame also works for other implementations of BST insert, other operations of BST, and other data structures such as linked lists.

\item[\S\ref{sec:tactic}] We demonstrate our tactic program for proving partial data structure's elimination rules and composition rules. 

\item[\S\ref{sec:comp}] We compare our proofs with traditional approaches which use recursively defined predicates. We discuss the power and limitation of using magic wand and we explain the name of our proof technique \emph{magic-wand-as-frame}.

%\item[\S 5] The proof in \S 2 also follows some other principles of proving imperative programs correct: ``splitting correctness proof into algorithm correctness proof and implementation correctness proof'' and ``isolate program variables in assertions''. We show that magic-wand-as-frame does still work with different choices on verification principles.

\item[\S\ref{sec:concl}] We discuss related work of using magic wand and summarize our contributions.

\end{itemize}

Remark 1. The purpose of this paper is NOT about GENERAL magic-wand-involved program verification. What we do propose is a very disciplined way of using magic wand, which can make program correctness proofs more elegant.

Remark 2. Using magic wand, using quantifiers and using high order logics are known methods and tools in program verification.
We do not claim the invention of any one of them.
Instead, we show a specific way to put them together and make proofs easier.

\section{Proof Pearl: Binary search tree insertion} \label{sec:paper_proof}
This section demonstrates the main content of this paper: a magic-wand-as-frame verification of BST insert.
Here we use standard mathematical notation;
in the next section we give details about the
Coq formalization and the proof notation of Verifiable C.

\subsection{Specification} \label{highlevelspec}

Correctness for BSTs means that the \lstinline{insert} function---considered as an operation of an
abstract data type---implements the \emph{update} operation
on finite maps from the key type (in this case, integer) to
the range type (in this case \lstinline{void*}).
The client of a \emph{finite map} does not need to know that trees
are used; we should hide that information in our specifications.
For that purpose, we define separation logic predicates for binary trees, and we define map predicates based on tree predicates. Only map predicates show up in the specification of this insert function.

\begin{figure}[h]
\vspace{-2ex}
\begin{align}
&\ \ \text{Binary trees: ~~~~~~~  $t =~~ E ~\mid~ T(t_1, k, v, t_2)$}   \notag \\
&  \notag \\
&\begin{array}{l}
 \mathrm{treebox\_rep}(E,p)~~\defeq~~ p \mapsto \mathrm{null}  \\
  \ \ \\
     \mathrm{treebox\_rep}(T(t_1, k, v, t_2),p)~~\defeq~~ \exists q. \\
      \ \ \ \ \ \ p \mapsto q \ * \ \fieldadd{q}{\mathsf{key}} \mapsto k  \ * \ \fieldadd{q}{\mathsf{value}} \mapsto v \ *\\
      \ \ \ \ \ \ \mathrm{treebox\_rep}(t_1, \fieldadd{q}{\mathsf{left}}) \ * \ \mathrm{treebox\_rep}(t_2, \fieldadd{q}{\mathsf{right}})
    \end{array} \label{eqn:treebox_rep} \\
&  \notag \\
&\begin{array}{l}
   \mathrm{Mapbox\_rep}(m, p)~~\defeq~~\exists t. \ \mathrm{Abs}(t, m) \wedge \mathrm{SearchTree}(t) \wedge \mathrm{treebox\_rep}(t,p)
   \end{array} \notag
\end{align}
\end{figure}

Here $\mathrm{SearchTree}(t)$ represents the \emph{search-tree property}. That is, at any node of the tree $t$,
the keys in the left subtree are strictly less than the key at
the node, and the keys in the right subtree are strictly greater.
$\mathrm{Abs}(t, m)$ is an abstraction relation, which says that $m$ is an abstraction of tree $t$, i.e., the key-value pairs in map $m$ and tree $t$ are identical.
$\mathrm{SearchTree}(t)$ and $\mathrm{Abs}(t, m)$ are formally defined in the SearchTree chapter of \emph{Verified Functional Algorithms} \cite{appel17:vfa}. Their exact definitions are not needed in our proof here.

Our high-level separation-logic specification of \lstinline{insert} function is:
\begin{align*}
\mathrm{Precondition:}~~~~& \{\denote{\mathsf{p}} = p_0 \wedge \denote{\mathsf{x}} = x \wedge  \  \denote{\mathsf{v}}=v \wedge \mathrm{Mapbox\_rep}(m_0, p_0) \} \\
\mathrm{Command:}~~~~& ~~~~ \mathsf{insert(p, x, v)} \\
\mathrm{Postcondition:}~~~~& \{\mathrm{Mapbox\_rep}(\mathrm{update}(m_0, x, v), p_0)\}
\end{align*}

We use $\denote{\mathsf{p}}$ to represent the value of program variable $\mathsf{p}$.
In the postcondition, $\mathrm{update}(m_0, x, v)$ is the usual update operation on maps.

\subsection{Two-level proof strategy}

One could directly prove the correctness
of the C-language \lstinline{insert} function, using
the \emph{search-tree} property as an invariant.
But it is more modular and scalable to do a two-level proof instead
%\cite{chargueraud2016higher}
\cite{appel16:modsec,gu15:popl}:
First, prove that the C program (imperatively, destructively) implements
the (mathematical, functional) $\ins$ function on binary search trees;
then prove that the (pure functional) binary search trees implement
(mathematical) finite maps, that $\ins$ implements update, and that $\ins$
preserves the search-tree property.
%To achieve such proofs in Coq, the set of all binary trees can be formalized in Coq as an inductive type and the mathmatical function $\ins$ can be formalized as a Coq function as shown at the beginning of this subsection.
So let us define insertion on pure-functional tree structures:
%from which the Gallina \lstinline{insert} function is taken. [Adobe illustrater]
\begin{align*}
\ins(E,x,v)~~\defeq~~ & T(E, x, v, E)  \\
\ins(T(t_1, x_0, v_0, t_2) ,x,v)~~\defeq~~ &
               \text{If } x < x_0, ~ T(\ins(t_1, x, v), x_0, v_0, t_2) \\
&           \text{If } x = x_0, ~ T(t_1, x, v, t_2) \\
&           \text{If } x > x_0, ~ T(t_1, x_0, v_0,\ins(t_2, x, v))
\end{align*}
The \emph{SearchTree} chapter of \emph{VFA} \cite{appel17:vfa}
proves (via the Abs relation) that this implements update on abstract
finite maps.  Next, we'll prove that the C program refines
this functional program; then compose the two proofs
to show that the C program satisfies its specification
given at the end of \S\ref{highlevelspec}.

For that refinement proof, we give a 
low-level separation-logic specification of the \lstinline{insert} function,
i.e., the C program refines the functional program:
\begin{align}
\mathrm{Precondition:}~~~~&
 \left\{\denote{\mathsf{p}} = p_0 \wedge \denote{\mathsf{x}} = x \wedge \denote{\mathsf{v}}=v \wedge \mathrm{treebox\_rep}(t_0, p_0) \right\} \notag \\
\mathrm{Command:}~~~~& ~~~~  \mathsf{insert(p, x, v)}  \label{eqn:implement_spec}\\
\mathrm{Postcondition:}~~~~& \{\mathrm{treebox\_rep}(\ins(t_0,x,v), p_0)\} \notag
\end{align}

\subsection{Magic wand for partial trees}

The function body of $\mathsf{insert}$ is just one loop.
We will need a loop invariant!
As shown in Fig. \ref{fig-exec}, the original binary tree can always be divided into two parts after every loop body iteration:
one is a subtree $t$ whose root is tracked by program variable $\mathsf{p}$ (that is, $\denote{\mathsf{*p}}$ is the address of $t$'s root node) and
another part is a partial tree $P$ whose root is identical with the original tree and whose hole is marked by address $\denote{\mathsf{p}}$.

The separation logic predicate for trees (also subtrees) is $\mathrm{treebox\_rep}$. We define the separation logic predicate for partial trees as follows.
Given a partial tree $P$, which is a function from binary trees to binary trees:
\begin{align*}
& \mathrm{partial\_treebox\_rep}(P, r, i) \defeq \forall t.  \left(\mathrm{treebox\_rep}(t, i) \wand \mathrm{treebox\_rep}(P(t), r)\right)
\end{align*}

This predicate has some important properties and we will use these properties in the verification of $\mathsf{insert}$.
(\ref{eqn:single_l}) and (\ref{eqn:single_r}) show how single-layer partial trees are constructed.
(\ref{eqn:emp_ptr}) shows the construction of empty partial trees. (\ref{eqn:tr_ptr}) shows that a subtree can be filled in the hole of a partial tree. And (\ref{eqn:ptr_ptr}) shows the composition of partial trees.

\begin{subequations}
\begin{align}
& \begin{array}{l} p \mapsto q \ * \ \fieldadd{q}{\mathsf{key}} \mapsto k \ * \ \fieldadd{q}{\mathsf{value}} \mapsto v \ * \ \mathrm{treebox\_rep}(t_2, \fieldadd{q}{\mathsf{right}}) \\ 
                            ~~~~ \vdash \mathrm{partial\_treebox\_rep}(\lambda t. \ T(t, k, v, t_2),~ p,~\fieldadd{q}{\mathsf{left}})
   \end{array} \label{eqn:single_l} \\
& \begin{array}{l} p \mapsto q \ * \ \fieldadd{q}{\mathsf{key}} \mapsto k \ * \ \fieldadd{q}{\mathsf{value}} \mapsto v \ * \ \mathrm{treebox\_rep}(t_1, \fieldadd{q}{\mathsf{left}}) \\ 
                            ~~~~ \vdash \mathrm{partial\_treebox\_rep}(\lambda t. \ T(t_1, k, v, t),~ p,~ \fieldadd{q}{\mathsf{right}})
   \end{array} \label{eqn:single_r}  \\
& \mathrm{emp} \vdash  \mathrm{partial\_treebox\_rep}(\lambda t. \ t,~ p,~ p) \label{eqn:emp_ptr} \\
& \begin{array}{l} \mathrm{treebox\_rep}(t, i) * \mathrm{partial\_treebox\_rep}(P, r, i) \\ 
                            ~~~~ \vdash  \mathrm{treebox\_rep}(P(t), r)
   \end{array} \label{eqn:tr_ptr} \\
& \begin{array}{l} \mathrm{partial\_treebox\_rep}(P_1,~ p_1,~ p_2) * \mathrm{partial\_treebox\_rep}(P_2,~ p_2,~ p_3) \\
                            ~~~~ \vdash  \mathrm{partial\_treebox\_rep}(P_1 \circ P_2,~ p_1,~ p_3)
   \end{array} \label{eqn:ptr_ptr}
\end{align}
\end{subequations}

\subsection{Wand-frame proof rules}

These properties are direct instances of the
\textsc{magic-wand-as-frame} proof rules,
which are all derived rules from minimum first-order separation logic (i.e. intuitionistic first order logic + commutativity and associativity of separating conjunction + $\mathrm{emp}$ being separating conjunction unit + \textsc{wand-ajoint}).

Here are the proof rules:
\begin{align*}
\hspace{-.1em}&\textsc{wandQ-frame-intro:} \\ & \ \ \ \ \ Q \vdash \forall x. \left( P(x) \wand P(x) * Q \right)\\ 
\hspace{-.1em}&\textsc{wandQ-frame-elim:} \\ & \ \ \ \ \ P(x) * \forall x. \left( P(x) \wand Q(x)\right) \vdash Q(x) \\
\hspace{-.1em}&\textsc{wandQ-frame-hor:} \\ & \ \ \ \ \  \forall x. \left( P_1(x) \wand Q_1(x)\right) * \forall x. \left( P_2(x) \wand Q_2(x)\right) \vdash \ \forall x. \left( P_1(x) * P_2(x) \wand Q_1(x) * Q_2(x)\right) \\
\hspace{-.1em}&\textsc{wandQ-frame-ver:} \\ & \ \ \ \ \  \forall x. \left( P(x) \!\wand \! Q(x)\right) * \forall x. \left( Q(x) \! \wand \! R(x) \right) \vdash  \  \forall x. \left( P(x) \! \wand R(x)\! \right) \\
\hspace{-.1em}&\textsc{wandQ-frame-refine:}\\ & \ \ \ \   \forall x. \left( P(x) \wand Q(x) \right) \vdash \forall y. \left( P(f(y)) \wand Q(f(y)) \right)
\end{align*}
%TODO really need refinement?
\textsc{wandQ-frame-intro} proves (\ref{eqn:single_l}), (\ref{eqn:single_r}) and (\ref{eqn:emp_ptr}). \textsc{wandQ-frame-elim} proves (\ref{eqn:tr_ptr}). \textsc{wandQ-frame-ver} and \textsc{wandQ-frame-refine} together prove (\ref{eqn:ptr_ptr}).

\emph{Remark 1.} Theoretically, in order to prove (\ref{eqn:single_l})-(\ref{eqn:ptr_ptr}), these \textsc{wandQ-frame} rules are the only additional proof rules that we need for magic wand and quantifiers.
But in practice, when we formalize these proofs in Coq,
we also use other proof rules like \textsc{wand-adjoint}
and the universal quantifier introduction rule.

\emph{Remark 2.} The soundness of (\ref{eqn:emp_ptr}) (\ref{eqn:tr_ptr}) and (\ref{eqn:ptr_ptr}) do not even depend on the definition of $\mathrm{treebox\_rep}$.

\subsection{Implementation correctness proof}
\label{loop-inv}

Now, we can verify the $\mathsf{insert}$ function with the loop invariant,
\[
\begin{array}{l}
\exists \ t \ p \ P. \ P(\mathrm{ins}(t, x, v)) = \mathrm{ins}(t_0, x, v) \ \wedge \ \denote{\mathsf{p}} = p \ \wedge \ \denote{\mathsf{x}} = x \ \wedge \ \denote{\mathsf{v}}=v  \ \wedge \\
\ \ \ \ \ \ \ \ \ \ \ \ \ \ \ \mathrm{treebox\_rep}(t, p)  \ * \ \mathrm{partial\_treebox\_rep}(P, p_0, p) \end{array}\]
It says, a partial tree $P$ and a tree $t$ are stored in disjoint pieces of memory, and if we apply the $\mathrm{ins}$ function to $t$ locally and fill the hole in $P$ with that result, then we will get the same as directly applying $\mathrm{ins}$ to the original binary tree $t_0$.

The correctness of $\mathsf{insert}$ is based on the following two facts.
First, the precondition of $\mathsf{insert}$
%$$\denote{\mathsf{p}} = p_0 \wedge \denote{\mathsf{x}} = x \wedge \denote{\mathsf{v}}=v \wedge \mathrm{treebox\_rep}(t_0, p_0)$$
implies this loop invariant because we can instantiate the existential variables $t$, $p$ and $P$ with $t_0$, $p_0$ and $\lambda t. \ t$ and apply property (\ref{eqn:emp_ptr}).
Second, the loop body preserves this loop invariant and every \lstinline|return| command satisfies the postcondition of the whole C function.
Fig. \ref{fig:loopbody} shows our proof (for conciseness, we omit $\denote{\mathsf{x}} = x \ \wedge \ \denote{\mathsf{v}}=v$ in all assertions).

\begin{figure*}
\begin{lstlisting}[language=C, numbers=left, numberstyle=\tiny, numbersep=4pt, 
    basicstyle=\bfseries, escapechar=|, backgroundcolor = \color{white}]
\\$\left\{\!\!\!\begin{array}{c@{}}\ P(\mathrm{ins}(t, x, v)) = \mathrm{ins}(t_0, x, v) \ \wedge \ \denote{\mathsf{p}} = p \ \wedge \\ \ \mathrm{treebox\_rep}(t, p) \ * \ \mathrm{partial\_treebox\_rep}(P, p_0, p)\end{array}\right\}$
q = * p;
if (q == NULL) {
   \\$\left\{\!\!\!\begin{array}{c@{}}\ P(\mathrm{ins}(t, x, v)) = \mathrm{ins}(t_0, x, v) \ \wedge \  \denote{\mathsf{p}} = p \ \wedge \ \denote{\mathsf{q}} = \mathrm{null} \ \wedge \ t = E \ \wedge \\ \ p \mapsto \mathrm{null} \ * \ \mathrm{partial\_treebox\_rep}(P, p_0, p) \end{array}\right\}$
   \\$\left\{\!\!\!\begin{array}{c@{}}\ P(T(E, x, v, E)) = \mathrm{ins}(t_0, x, v) \ \wedge \ \denote{\mathsf{p}} = p \ \wedge \\ \ p \mapsto \mathrm{null} \ * \ \mathrm{partial\_treebox\_rep}(P, p_0, p)  \end{array}\right\}$
   q = (struct tree *) surely_malloc (sizeof *q);
   q$\to$key=x; q$\to$value=v;
   q$\to$left=NULL; q$\to$right=NULL;
   *p=q;
   \\$\left\{\!\!\!\begin{array}{c@{}}\ P(T(E, x, v, E)) = \mathrm{ins}(t_0, x, v) \ \wedge \ \denote{\mathsf{p}} = p \ \wedge \\ \ \mathrm{treebox\_rep}(T(E, x, v, E), p) \ * \ \mathrm{partial\_treebox\_rep}(P, p_0, p)\end{array}\right\}$ |\label{line:case1-exit1-pre}|
   \\$\{P(T(E, x, v, E)) = \mathrm{ins}(t_0, x, v) \ \wedge \ \denote{\mathsf{p}} = p \ \wedge \ \mathrm{treebox\_rep}(P(T(E,x,v,E)), p_0)\}$ |\label{line:case1-exit1-post}|
   \\$\{\mathrm{treebox\_rep}(\ins(t_0, x, v), p_0)\}$
   return; |\label{line:case1-exit1}|
} else {
   \\$\left\{\!\!\!\begin{array}{l@{}}\ \exists t_1 \ t_2 \ x_0 \ v_0 \ q. \ \ \ P(\mathrm{ins}(t, x, v)) = \mathrm{ins}(t_0, x, v) \ \wedge \ t = T(t_1, x_0, v_0, t_2)  \ \wedge \ \denote{\mathsf{p}} = p \ \wedge \\ \ \denote{\mathsf{q}} = q \ \wedge \ p \mapsto q \ * \ \fieldadd{q}{\mathsf{key}} \mapsto x_0 \ * \ \fieldadd{q}{\mathsf{value}} \mapsto v_0  \ * \ \mathrm{treebox\_rep}(t_1, \fieldadd{q}{\mathsf{left}}) \ * \\ \ \mathrm{treebox\_rep}(t_2, \fieldadd{q}{\mathsf{right}}) \ * \ \mathrm{partial\_treebox\_rep}(P, p_0, p)  \end{array}\right\}$ |\label{line:case1-else-pre}|
    int y = q$\to$key; $\label{line:assign-cmd-234}$
    if (x<y)
        p= &q$\to$left; |\label{line:case1-exit2}|
        \\$\left\{\!\!\!\begin{array}{l@{}}\ \exists t_1 \ t_2 \ x_0 \ v_0 \ q. \ \ \ x < x_0 \ \wedge \ P(\mathrm{ins}(t, x, v)) = \mathrm{ins}(t_0, x, v) \ \wedge \\ \ t = T(t_1, x_0, v_0, t_2)  \ \wedge \ \denote{\mathsf{p}} = \fieldadd{q}{\mathsf{left}} \ \wedge \\ \ \mathrm{partial\_treebox\_rep}(\lambda \hat{t}. \ T(\hat{t}, x_0, v_0, t_2), p, \fieldadd{q}{\mathsf{left}}) \ * \\ \ \mathrm{treebox\_rep}(t_1, \fieldadd{q}{\mathsf{left}}) \ * \ \mathrm{partial\_treebox\_rep}(P, p_0, p)\end{array}\right\}$ |\label{line:case1-exit2-pre}|
        \\$\left\{\!\!\!\begin{array}{l@{}}\ \exists t_1 \ t_2 \ x_0 \ v_0 \ q. \ \ \ P(T(\mathrm{ins}(t_1, x, v), x_0, v_0, t_2)) = \mathrm{ins}(t_0, x, v) \ \wedge \\ \ \denote{\mathsf{p}} = \fieldadd{q}{\mathsf{left}} \ \wedge \ \mathrm{treebox\_rep}(t_1, \fieldadd{q}{\mathsf{left}}) \ * \\ \ \mathrm{partial\_treebox\_rep}(\lambda \hat{t}. \ P(T(\hat{t}, x_0, v_0, t_2)), p_0, \fieldadd{q}{\mathsf{left}}) \end{array}\right\} \label{line:case1-exit2-post}$
    else if (y<x)
        p= &q$\to$right; |\label{line:case1-exit3}|
        \\$\left\{\!\!\!\begin{array}{l@{}}\ \exists t_1 \ t_2 \ x_0 \ v_0 \ q. \ \ \ P(t_1, x_0, v_0, T(\mathrm{ins}(t_2, x, v))) = \mathrm{ins}(t_0, x, v) \ \wedge \\ \ \denote{\mathsf{p}} = \fieldadd{q}{\mathsf{right}} \ \wedge \ \mathrm{treebox\_rep}(t_2, \fieldadd{q}{\mathsf{right}}) \ * \\ \ \mathrm{partial\_treebox\_rep}(\lambda \hat{t}. \ P(T(t_1, x_0, v_0, \hat{t})), p_0, \fieldadd{q}{\mathsf{right}}) \end{array}\right\}$
    else { 
        p$\to$value=v; |\label{line:case1-exit4}|
        \\$\left\{\!\!\!\begin{array}{l@{}}\ \exists t_1 \ t_2 \ x_0 \ v_0 \ q. \ \ \ x = x_0 \ \wedge \ P(\mathrm{ins}(t, x, v)) = \mathrm{ins}(t_0, x, v) \ \wedge \ t = T(t_1, x_0, v_0, t_2)  \ \wedge \\ \mathrm{treebox\_rep}(T(t_1, x, v, t_2), p) \  * \ \mathrm{partial\_treebox\_rep}(P, p_0, p) \end{array}\right\}$
        \\$\left\{\!\!\!\begin{array}{l@{}}\ \exists t_1 \ t_2 \ x_0 \ v_0 \ q. \ \ \ P(T(t_1, x, v, t_2)) = \mathrm{ins}(t_0, x, v) \ \wedge \\ \  \mathrm{treebox\_rep}(T(t_1, x, v, t_2), p) \  * \ \mathrm{partial\_treebox\_rep}(P, p_0, p) \end{array}\right\}$
        \\$\{\mathrm{treebox\_rep}(\ins(t_0, x, v), p_0)\}$
        return; |\label{line:case1-exit4-post}|
   } }
\end{lstlisting}
\vspace{-2ex}
\caption{Proof of loop body}
\label{fig:loopbody}
\end{figure*}

This loop body has four branches: two of them end with return commands and the other two end normally.
In the first branch, the inserted key does not appear in the original tree. This branch ends with a \lstinline|return| command at line \ref{line:case1-exit1}.
We show that the program state at that point satisfies the postcondition of the whole function body (line \ref{line:case1-exit1-post}).
The transition from line \ref{line:case1-exit1-pre} to line \ref{line:case1-exit1-post} is sound due to rule (\ref{eqn:tr_ptr}).
The second branch contains only one command at line \ref{line:case1-exit2}. We re-establish our loop invariant in this branch (line \ref{line:case1-exit2-post}).
The transition from line \ref{line:case1-else-pre} to line \ref{line:case1-exit2-pre} is due to rule (\ref{eqn:single_l}) and the transition from line \ref{line:case1-exit2-pre} to line \ref{line:case1-exit2-post} is due to rule (\ref{eqn:ptr_ptr}).
The third branch at line \ref{line:case1-exit3} is like the second,
and the last branch is like the first one.

$~~~$\includegraphics[scale=2]{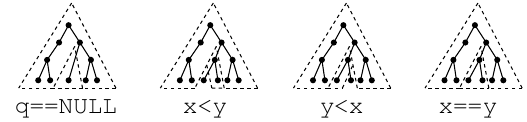}

In summary, the partial tree $P$ is established as an empty partial tree ($\lambda \hat{t}. \ \hat{t}$) in the beginning.
The program merges one small piece of subtree $t$ into the partial tree in each iteration of the loop body.
Finally, when the program returns, it establishes a local insertion result ($\mathrm{ins}(t, x, v)$) and fills it in the hole of that partial tree---we know the result must be equivalent with directly applying insertion to the original binary tree. The diagrams above illustrate the situations of these four branches and our proof verifies this process.

\section{Coq formalization in Verifiable C} \label{sec:Coq_proof}
We machine-check this proof in Coq,
using the Verified Software Toolchain's \emph{Verifiable C} program logic \cite{appel14:plcc},
which is already proved sound w.r.t. CompCert Clight \cite{blazy2009:clight}.
We import from \emph{Verified Functional Algorithms}
the definition of purely functional search trees and their properties.
Readers can find our Coq development online: 

\lstinline|https://github.com/PrincetonUniversity/VST/tree/master/wand_demo|

We formalize our proof using Verifiable C's
interactive symbolic execution system in Coq \cite{cao18:floyd}.
Until now, Verifiable C had not included much proof theory for wand,
except the basic \textsc{wand-adjoint}.  Now we add
the \textsc{proof rules of wand-frame} (see \lstinline|wandQ_frame.v|) as derived lemmas from Verifiable C's basic separation logic.
We use them in the Coq proof of $\mathrm{partial\_treebox\_rep}$'s properties (see \S\ref{sec:PTR} and \lstinline|bst_lemmas.v|).

\subsection{Separation logic predicates and properties for BST}\label{sec:PTR}

Binary trees with keys and values are already formalized in VFA
as an inductive data type in Coq.
Here, we will formalize the separation logic predicate $\mathrm{treebox\_rep}$.
Instead of defining $\mathrm{treebox\_rep}$ directly as in (\ref{eqn:treebox_rep}), we first define $\mathrm{tree\_rep}$, then define $\mathrm{treebox\_rep}$ based on that.
Finally, we prove that it satisfies the equalities in (\ref{eqn:treebox_rep}).
We choose to do this because C functions for BST operations do not always take arguments with type \lstinline|(struct tree * *)| (or equivalently, \lstinline|treebox|).
For example, a look-up operation does not modify a BST, so it can just take a BST by an argument with type \lstinline|(struct tree *)|.

\begin{lstlisting}
  Fixpoint tree_rep (t: tree val) (p: val) : mpred :=
   match t with
   | E => !!(p=nullval) && emp
   | T a x v b =>
      EX pa:val, EX pb:val,
         data_at Tsh t_struct_tree (Vint (Int.repr (Z.of_nat x)),(v,(pa,pb))) p *
         tree_rep a pa * tree_rep b pb
   end.

  Definition treebox_rep (t: tree val) (b: val) :=
   EX p: val, data_at Tsh (tptr t_struct_tree) p b * tree_rep t p.
\end{lstlisting}

\pagebreak
\begin{lstlisting}
  Lemma treebox_rep_spec: forall (t: tree val) (b: val),
    treebox_rep t b =
    EX p: val,
    data_at Tsh (tptr t_struct_tree) p b *
    match t with
    | E => !!(p=nullval) && emp
    | T l x v r =>
        field_at Tsh t_struct_tree [StructField _key] (Vint (Int.repr (Z.of_nat x))) p *
        field_at Tsh t_struct_tree [StructField _value] v p *
        treebox_rep l (field_address t_struct_tree [StructField _left] p) *
        treebox_rep r (field_address t_struct_tree [StructField _right] p)
    end.
\end{lstlisting}

Here, \lstinline|val| is CompCert Clight's value type; \lstinline|nullval| has type \lstinline{val} and represents the value of \lstinline{NULL} pointer. The Coq type \lstinline{mpred} is the type of Verifiable C's separation logic predicates. ``\lstinline|&&|'', ``$*$'' and ``$\mathrm{EX}$'' are notations for conjunction, separating conjunction and existential quantifiers in Verifiable C's assertion language. ``\lstinline{!! _}'' is the notation that injects Coq propositions into the assertion language. The expression \lstinline|(Vint (Int.repr (Z.of_nat x))|  injects a natural number \lstinline|x| into the integers, then to a 32-bit integer,\footnote{Mapping $\mathbb{Z}$ to $\mathbb{Z}\!\mod 2^{32}$ is not injective; in a practical application the client
of this search-tree module should prove that $x<2^{32}$.}
 then to CompCert Clight's value type, \lstinline|val|.

\lstinline{Data_at} is a mapsto-like predicate for C aggregate types.  Here,
\begin{lstlisting}
  data_at Tsh t_struct_tree (Vint (Int.repr (Z.of_nat x)),(v,(pa,pb))) p
\end{lstlisting}
means that \lstinline|x, v, pa, pb| are four fields of the ``\lstinline|struct tree|'' stored at address \lstinline|p|. \lstinline|Tsh| means top share (full read/write permission).
Verifiable C's
\lstinline|field_at| is like \lstinline|data_at| but
permits a field name such as ``$\fieldadd{}{\mathsf{right}}$''.

Our partial tree predicate
\lstinline{partialT} \emph{does not care} how \lstinline{treebox_rep} works
internally.  Thus, in defining the proof theory of partial trees,
we parameterize over the treebox predicate.
In consequence, these parameterized proofs can be applied on both \lstinline{partial_treebox_rep} and \lstinline{partial_tree_rep}.
\begin{lstlisting}
Definition partialT (rep: tree val $\to$ val $\to$ mpred) (P: tree val $\to$ tree val) (p_root p_in: val) :=
      ALL t: tree val, rep t p_in $\wand$ rep (P t) p_root.
Definition partial_treebox_rep := partialT treebox_rep.
Definition partial_tree_rep := partialT tree_rep.
\end{lstlisting}
As claimed in \S\ref{sec:paper_proof}, the soundness of rules (\ref{eqn:emp_ptr}) (\ref{eqn:tr_ptr}) and (\ref{eqn:ptr_ptr}) do not depend on the definition of $\mathrm{treebox\_rep}$; we prove them sound for arbitrary partial tree predicates. For example, the following lemma is the generalized version of (\ref{eqn:tr_ptr}). We directly prove it by \textsc{wandQ-frame-elim}.
\begin{lstlisting}
Lemma rep_partialT_rep: forall rep t P p q, rep t p * partialT rep P q p $\vdash$ rep (P t) q.
Proof. intros. exact (wandQ_frame_elim _ (fun t => rep t p) (fun t => rep (P t) q) t). Qed.
\end{lstlisting}
We define $\mathrm{Mapbox\_rep}$ based on $\mathrm{treebox\_rep}$, $\mathrm{Abs}$ and $\mathrm{SearchTree}$ as described in \S\ref{sec:paper_proof};
and $\mathrm{Abs}$ and $\mathrm{SearchTree}$ are already defined in VFA. Similarly, we define $\mathrm{Map\_rep}$ based on $\mathrm{tree\_rep}$; application of it can be found in \S\ref{sec:other_proof}.

\subsection{C program specification and verification}

\emph{Specification and Coq proof goal.}
Verifiable C requires users to write C function specification in a canonical form.
The following is the specification of C function $\mathrm{insert}$. The \lstinline|WITH| clause there says that this specification is a parameterized Hoare triple---that is, for any \lstinline|p0, x, v, m0|, this specific triple is valid.
The brackets after \lstinline|PRE| hold the C argument list. CompCert Clight turns every C variable into an identifier in the Clight abstract syntax tree defined in Coq.
In this argument list, \lstinline|_p| is the identifier for C variable $\mathsf{p}$, etc. The brackets after \lstinline|POST| hold the C function return type.

\begin{lstlisting}
Definition insert_spec :=
 DECLARE _insert
  WITH p0: val, x: nat, v: val, m0: total_map val
  PRE  [ _p OF (tptr (tptr t_struct_tree)), _x OF tint, _v OF (tptr Tvoid)   ]
    PROP()
    LOCAL(temp _p p0; temp _x (Vint (Int.repr (Z.of_nat x))); temp _v v)
    SEP (Mapbox_rep m0 p0)
  POST [ Tvoid ] 
    PROP() LOCAL() SEP (Mapbox_rep (t_update m0 x v) p0).
\end{lstlisting}

Both precondition and postcondition are written in a  \lstinline|PROP/LOCAL/SEP| form. \lstinline|PROP| clauses are for program-variable-irrelevant pure facts;
there happen to be none here. \lstinline|LOCAL| clauses talk about the values of program variables. For example, \lstinline|temp _p p0| says $\denote{\mathsf{p}} = p_0$.
\lstinline|SEP| clauses are separating conjuncts. Verifiable C requires users to isolate programs variables in their assertions---\lstinline{SEP} conjuncts
do not refer directly to C program variables---so we use \lstinline|LOCAL| clauses to connect program variables to \lstinline{PROP} and \lstinline{SEP} clauses.

\vspace{\baselineskip}

\begin{lstlisting}
Theorem body_insert: semax_body Vprog Gprog f_insert insert_spec.
 (* The C function f_insert (Fig. 1) implements its specification *)
Proof.
  start_function.
  . . . (* 6 lines of reduction proof and *)
  . . . (* 65 lines of forward proof in separation logic *)
Qed.  
\end{lstlisting}

The proof contains two parts. One is to reduce the abstract specification to the concrete specification (6 lines).
The other is forward verification using separation logic (65
 lines)---it shares exactly the same structure with the pen-and-paper proof (30 lines) in Fig. \ref{fig:loopbody}.
The proof scripts can be found in our Coq development. We omit them here.

\section{Other data structures, programs and proofs} \label{sec:other_proof}

Magic-wand-as-frame is a pretty flexible proof technique. We briefly introduce some other possibilities in magic-wand-as-frame proofs here.
Interested readers can download our Coq development for more details.

%\paragraph{Alternative magic-wand-as-frame proofs for BST insert.} Universal quantifiers are not necessary for magic-wand-as-frame proofs.
%In the BST \lstinline{insert} example, we can also use  
%\[
%\begin{array}{l}
%\exists \ t \ p . \ \denote{\mathsf{p}} = p \ \wedge \ \denote{\mathsf{x}} = x \ \wedge \ \denote{\mathsf{v}}=v  \ \wedge \ \mathrm{treebox\_rep}(t, p)  \ * \\
%\ \ (\mathrm{treebox\_rep}(\mathrm{ins}(t, x, v), p) \wand \mathrm{treebox\_rep}(\mathrm{ins}(t_0, x, v), p_0)) \end{array}\]
%as loop invariant. The proofs are very similar except that we can use \textsc{wand-frame} rules instead of \textsc{wandQ-frame} rules to generate properties of partial tree predicates.
%\begin{align*}
%&\textsc{Quantifier-free proof rules of wand-frame}\mathsf{(wand\_frame.v)}:\\
%&\textsc{wand-frame-intro: } ~~ Q \vdash P \wand P * Q \\
%&\textsc{wand-frame-elim: } ~~ P * (P \wand Q) \vdash Q \\
%&\textsc{wand-frame-ver: } ~~ (P \wand Q) * (Q \wand R) \vdash P \wand R \\
%&\textsc{wand-frame-hor: } ~~ (P_1 \wand Q_1) * (P_2 \wand Q_2) \vdash P_1 * P_2 \wand Q_1 * Q_2
%\end{align*}

\subsection{Other BST operations.} We also verify C implementations of BST delete and look-up operation with the magic-wand-as-frame technique. In the verification of BST delete, we also use $\mathrm{partial\_treebox\_rep}$ to describe partial trees and use rules (\ref{eqn:single_l}-\ref{eqn:ptr_ptr})  to complete the proof. In the verification of BST look-up, we define $\mathrm{partial\_tree\_rep}$ using parameterized \lstinline|partialT| (see \S\ref{sec:Coq_proof}) and prove similar proof rules for it.
Especifically, we get the counterparts of (\ref{eqn:emp_ptr}-\ref{eqn:ptr_ptr}) for free because we have already proved them for general \lstinline|partialT| predicates. Proofs of the other two are also very straightforward using \textsc{wandQ-intro}.

\subsection{Other data structure: linked list.}
We also use magic-wand-as-frame to verify linked list append (see \lstinline|verif_list.v|). In that proof, we use the following separation logic predicates and proof rules (see \lstinline|list_lemmas.v|). These proof rules are direct instances of \textsc{wandQ-frame} rules. Here, we use $l_1l_2$ to represent the list concatenation of $l_1$ and $l_2$.
\begin{align}
&\begin{array}{lcl}
p \listrep [] & \defeq & p = \mathrm{null}  \wedge \mathrm{emp}\\
p \listrep (h :: t) & \defeq&
\fieldadd{p}{\mathsf{head}} \mapsto h \ * \ \exists q. \ \fieldadd{p}{\mathsf{tail}} \mapsto q \ * \  q \listrep t \\
p \lseg{l} q & \defeq & \forall l'. \ \left( q \listrep l' \ \wand  \ p \listrep ll' \right)
\end{array} \notag \\[1ex]
&\fieldadd{p}{\mathsf{head}} \mapsto h \ * \ \fieldadd{p}{\mathsf{tail}} \mapsto q \ \vdash \ p \lseg{[h]} q \notag \\
& \mathrm{emp} \ \vdash \ p \lseg{[]} p \notag \\
&p \lseg{l_1} q \ * \ q \listrep l_2 \ \vdash \  p \listrep l_1l_2 \notag \\
& p \lseg{l_1} q \ * \ q \lseg{l_2} r \ \vdash \ p \lseg{l_1l_2} r \label{eqn:lseg_lseg}
\end{align}

\subsection{Other data structure: C aggregate type.}
Another application of magic-wand-as-frame to verify load/store rules for nested C structs. For example, the following is a nested struct definition in C.
\begin{lstlisting}[language=C]
struct s1 { int f11; int f12; int f13; int f14; int f15; };
struct s2 { struct s2 f21; struct s2 f22; struct s2 f23; };
struct s2 * p; int x;
\end{lstlisting}
We want to derive the following rules
\begin{align*}
& \{\denote{\mathsf{p}} \mapsto u\} \ \ \  \ \mathsf{x = (*p).f2i.f1j} \ \ \ \ \{\denote{\mathsf{x}}=u_{i,j} \wedge \denote{\mathsf{p}} \mapsto u\} \\
& \{\denote{\mathsf{x}} = w \ \wedge \ \denote{\mathsf{p}} \mapsto u\} \  \ \ \ \mathsf{(*p).f2i.f1j = x} \ \ \  \{\denote{\mathsf{p}} \mapsto u[i,j \to w]\}
\end{align*}
from the following primary rules:
\begin{align*}
& \{\denote{\mathsf{p}}\mathsf{.f2i.f1j} \mapsto u_{i,j} \}  \ \ \ \ \mathsf{x = (*p).f2i.f1j} \ \ \ \ \{\denote{\mathsf{x}}=u_{i,j}  \wedge \denote{\mathsf{p}}\mathsf{.f2i.f1j} \mapsto u_{i,j} \} \\
& \{\denote{\mathsf{x}} = w \ \wedge \ \denote{\mathsf{p}}\mathsf{.f2i.f1j} \mapsto u_{i,j}\} \ \ \ \mathsf{(*p).f2i.f1j = x} \ \ \ \ \{\denote{\mathsf{p}}\mathsf{.f2i.f1j} \mapsto w\}
\end{align*}
in which $u$ represents a tuple of integer tuples, $u_{i,j}$ represents the $j$-th element of the $i$-th element of $u$ and $u[i,j \to w]$ is the result of replacing the $j$-th element of the $i$-th element of $u$ with $w$. And later, we will use $u_i$ to represent the $i$-th element of $u$. We will use $u[i \to v]$ and $u_i[j \to w]$ to represent results of replacing the $i$-th element of $u$, and replacing the $j$-th element of $u_i$.

The derivation above is completed by rule of consequence, the frame rule and the following facts (\ref{eqn:loadstore_pre}) (\ref{eqn:loadstore_post}):
\begin{align}
& \denote{\mathsf{p}} \mapsto u \notag\\
 \vdash & \denote{\mathsf{p}}\mathsf{.f2i} \mapsto u_i \ * \  \forall v. (\denote{\mathsf{p}}\mathsf{.f2i} \mapsto v \ \wand \ \denote{\mathsf{p}} \mapsto u[i \to v]) \notag\\
 \vdash & \denote{\mathsf{p}}\mathsf{.f2i.f1j} \mapsto u_{i,j} \ * \notag\\
 & \ \ \  \forall w. (\denote{\mathsf{p}}\mathsf{.f2i.f1j} \mapsto w \ \wand \ \denote{\mathsf{p}}\mathsf{.f2i} \mapsto u_i[j \to w]) \ * \notag\\
 & \ \ \  \forall v. (\denote{\mathsf{p}}\mathsf{.f2i} \mapsto v \ \wand \ \denote{\mathsf{p}} \mapsto u[i \to v]) \notag\\
 \vdash & \denote{\mathsf{p}}\mathsf{.f2i.f1j} \mapsto u_{i,j} \ * \notag\\
 & \ \ \  \forall w. (\denote{\mathsf{p}}\mathsf{.f2i.f1j} \mapsto w \ \wand \ \denote{\mathsf{p}}\mathsf{.f2i} \mapsto u_i[j \to w]) \ * \notag\\
 & \ \ \  \forall w. (\denote{\mathsf{p}}\mathsf{.f2i} \mapsto u_i[j \to w] \ \wand \ \denote{\mathsf{p}} \mapsto u[i \to u_i[j \to w]]) \notag\\
 \vdash & \denote{\mathsf{p}}\mathsf{.f2i.f1j} \mapsto u_{i,j} \ * \notag\\
 & \ \ \  \forall w. (\denote{\mathsf{p}}\mathsf{.f2i.f1j} \mapsto w \ \wand \ \denote{\mathsf{p}}\mathsf{.f2i} \mapsto u_i[j \to w]) \ * \notag\\
 & \ \ \  \forall w. (\denote{\mathsf{p}}\mathsf{.f2i} \mapsto u_i[j \to w] \ \wand \ \denote{\mathsf{p}} \mapsto u[i,j \to w]) \notag\\
 \vdash & \denote{\mathsf{p}}\mathsf{.f2i.f1j} \mapsto u_{i,j} \ * \  \forall w. (\denote{\mathsf{p}}\mathsf{.f2i.f1j} \mapsto w \ \wand \ \denote{\mathsf{p}} \mapsto u[i,j \to w])  \label{eqn:loadstore_pre}
\end{align}
Here, the first two turnstiles are due to \textsc{wandQ-frame-intro}. The third turnstile is due to \textsc{wandQ-frame-refine}.
The forth turnstile is due to the fact that $$u[i \to u_i[j \to w]] = u[i,j \to w].$$ And the last turnstile is due to \textsc{wandQ-frame-ver}.
Moreover, \textsc{wandQ-frame-elim} gives us:
\begin{align}
 & \denote{\mathsf{p}}\mathsf{.f2i.f1j} \mapsto w \ * \  \forall w. (\denote{\mathsf{p}}\mathsf{.f2i.f1j} \mapsto w \ \wand \ \denote{\mathsf{p}} \mapsto u[i,j \to w]) \
\vdash  \denote{\mathsf{p}} \mapsto u[i,j \to w] \label{eqn:loadstore_post}
\end{align}

In Coq, we verify general proof rules for nested load/store for any nested C structs and any field paths by induction over the length of the paths and by our magic-wand-as-frame technique.

\section{Tactics for eliminating and/or merging wand-frame predicates}
\label{sec:tactic}
In \S\ref{sec:Coq_proof}, we prove the elimination rule of \lstinline{partial_treebox_box} (\lstinline{rep_partialT_rep}) using only one line of Coq proof. 
Obviously, there is a common pattern for verifying different wand frame predicates' elimination rules.
We design a Coq tactic for completing these proofs automatically. This tactic can even prove the following entailment which needs multiple elimination steps:
\[ p \lseg{l_1} q \ * \ q \lseg{l_2} r \ * \ r \listrep l_3 \ \vdash \  p \listrep l_1(l_2l_3). \]

Here is a brief description of tactic ``\lstinline{sep_absorb $P$}''.
It is designed to simplify a separation logic entailment with the following form (and prove it if possible).
\[P * \forall x. (Q_1(x) \wand R_1(x)) * \forall x. (Q_2(x) \wand R_2(x)) * \dots * \forall x. (Q_n(x) \wand R_n(x)) \vdash S.\]
\begin{itemize}
\item[] Step 1. Pick a quantified wand expression $\forall x. (Q_i(x) \wand R_i(x))$ on the left side.
\item[] Step 2. Let Coq try unifying $P$ and $Q_i(?x)$, in which $?x$ represents an evar in Coq.
\item[] Step 3. If succeeded, assuming $Q_i(?x)$ is instantiated as $Q_i(x_0) = P$, then on the left hand side replace $(P * \forall x. (Q_i(x) \wand R_i(x)))$ with $R_i(x_0)$. Call ``\lstinline{sep_absorb $R_i(x_0)$}'' recursively.
\item[] Step 4. If failed, repeat steps 1-3 for all other possible separting conjuncts on the left hand side.
\item[] Step 5. If all attempts failed, test whether the left hand side is identical with the right hand side. If yes, prove it by the reflexivity of entailment relation.
\end{itemize}

The separation logic transformation of step 3 is done by \lstinline{eapply} the following lemma. In this lemma, \lstinline|(ALL x: A, QR x)| represents $(\forall x. (Q_i(x) \wand R_i(x)))$ and \lstinline|QR'| represents other separating conjuncts. Full tactic definition can be found in \lstinline|wandQ_frame_tactic.v|.

\begin{lstlisting}
Lemma wandQ_elim_alg2: forall {A} P (QR: A $\to$ mpred) QR' RHS R x0,
QR x0 = P $\wand$ R x0 $\to$
R x0 * QR' $\vdash$ RHS $\to$
P * (ALL x: A, QR x) * QR' $\vdash$ RHS.
\end{lstlisting}

We also implement a tactic for proving vertical composition rules based on \lstinline{sep_absorb}. The main idea is that the following two claims are equivalent. We reduce the former one to the latter one.
\[\forall x. (Q_1(x) \wand R_1(x)) * \dots * \forall x. (Q_n(x) \wand R_n(x)) \vdash \forall x. (P(x) \wand S(x));\]
\[\text{For any } a, P(a) * \forall x. (Q_1(x) \wand R_1(x)) *  \dots * \forall x. (Q_n(x) \wand R_n(x)) \vdash S(a).\]

\section{Magic-wand-as-frame vs. traditional proofs} \label{sec:comp}
We have used magic wand to define partial tree (tree-with-a-hole) predicates
and list segment (list-with-a-hole) predicates.  Berdine \emph{et al.} \cite{berdine05}
first defined \emph{list segments} and demonstrated a proof of imperative list append;
Chargu\'{e}raud defined tree-with-holes for a proof of BST operations.

These authors defined partial tree (and also list segment)
by an explicit inductive definition,
roughly as follows:

\begin{align*}
& \text{Partial trees:} \ \  P = H ~\mid~ L(P, k, v, t_2) ~\mid~ R(t_1, k, v, P) \\
& \\
& \mathrm{partial\_treebox\_rep^R}(H,r,i)~~\defeq~~ \\
& \ \ \ \  r = i \wedge \mathrm{emp} \\
& \mathrm{partial\_treebox\_rep^R}(L (P, k, v, t_2), r, i)~~\defeq~~  \\
& \ \ \ \ \exists q. \ r \mapsto q \ * \ \fieldadd{q}{\mathsf{key}} \mapsto k  \ * \ \fieldadd{q}{\mathsf{value}} \mapsto v \ *\\
& \ \ \ \ \mathrm{partial\_treebox\_rep^R}(P, \fieldadd{q}{\mathsf{left}}, i) \ * \\
& \ \ \ \ \mathrm{treebox\_rep}(t_2, \fieldadd{q}{\mathsf{right}}) \\
& \mathrm{partial\_treebox\_rep^R}(R (t_1, k, v, P), r, i)~~\defeq~~ ... \\
\end{align*}

That is: a partial tree is either one single hole or a combination of a partial tree and a complete tree; the partial tree can act as either the left subtree or the right subtree. And $\mathrm{partial\_treebox\_rep^R}$ is defined as a \emph{recursive} predicate over that structure.

\subsection{Advantages of using wand}

With this alternative definition, proof rules (\ref{eqn:single_l})--(\ref{eqn:ptr_ptr}) are still sound and our proof in Fig. \ref{fig:loopbody} still holds. However, our magic wand approach is better than that in three aspects.

\emph{Parameterized definition and proofs.} Using magic wand, we can define \lstinline|partialT| as a parameterized predicate for partial trees and proof rules (\ref{eqn:emp_ptr})--(\ref{eqn:ptr_ptr}) are sound in that parameterized way. Both $\mathrm{partial\_treebox\_rep}$ and $\mathrm{partial\_tree\_rep}$ are its instances.

\emph{Domain-specific theories for free.} When a partial tree is defined as a function from trees to trees, we get the definition of ``filling the hole in $P$ with tree $t$'' and ``shrinking the hole in $P_1$ with another partial tree $P_2$'' for free. They are just $P(t)$ and $P_1 \circ P_2$. Also, the following equations are for free:
$$(P_1 \circ P_2)(t) = P_1 (P_2 (t)) \quad \quad \quad P_1 \circ (P_2 \circ P_3) = (P_1 \circ P_2) \circ P_3$$

It is not obvious that we can formalize partial trees as functions from trees to trees! Because not all functions from trees to trees are partial trees.

In fact, if the predicate for partial trees were defined recursively (as Chargu\'{e}raud proposed), we would have to define partial trees as a Coq inductive type.
Thus, two combinators above must be defined Coq recursive functions and we would have to prove the equations above by induction.

\emph{Avoiding brittle and complex separation logic proofs.} Using magic wand and quantifiers, rule (\ref{eqn:tr_ptr}) and (\ref{eqn:ptr_ptr}) are direct corollaries of \textsc{wandQ-frame} rules. However, proving them from recursively defined $\mathrm{partial\_treebox\_rep^R}$ needs induction over the partial tree structure.

In some situation, these induction proofs can be very complicated and even annoying to formalize in Coq. The separation logic predicate for C aggregate types is such an example. The \lstinline|data_at| predicate is already dependently typed. Proof rules that substitute a single field's data are now described by magic-wand-involved expressions. Their soundness proofs are significantly shorter (although still quite long) than using hole-related predicates.

Even worse, those inductive proofs are actually very brittle beside their length and complexity. Using linked-list predicates as an example, different authors had proposed different recursive definitions for list segments. Here is Smallfoot's \cite{berdine05} definition:
\begin{align*}
& \ \ p \lseg{[]} q ~~\defeq~~ p = q  \wedge \mathrm{emp}\\
& \begin{array}{l}
p \lseg{(h :: t)} r ~~\defeq \exists q. \ p \neq r \ \wedge \ \fieldadd{p}{\mathsf{head}} \mapsto h \ * \ \fieldadd{p}{\mathsf{tail}} \mapsto q \ * \  q \lseg{t} r
\end{array} \\
& \ \ p \listrep l ~~\defeq~~ p \lseg{l} \mathrm{null}
\end{align*}
And here is the definition from  Chargu\'eraud \cite{chargueraud2016higher}:
\begin{align*}
& \ \ p \lseg{[]} q ~~\defeq~~ p = q  \wedge \mathrm{emp}\\
& \begin{array}{l}
p \lseg{(h :: t)} r ~~\defeq~~ \exists q. \ s\fieldadd{p}{\mathsf{head}} \mapsto h \ * \ \fieldadd{p}{\mathsf{tail}} \mapsto q \ * \  q \lseg{t} r
\end{array} \\
& \ \ p \listrep l ~~\defeq~~ p \lseg{l} \mathrm{null}
\end{align*}
These two definitions look similar, but their proof theories are surprisingly different. Proof rules in (\ref{eqn:lseg_lseg}) are unsound with respect to Smallfoot's definition but sound with respect to Chargu\'eraud's definition. Specifically, SmallFoot's list segment only satisfies weaker rules like the following one:
\begin{align*}
& p \lseg{l_1} q \ * \ q \lseg{l_2} r \ * \ r \listrep l_3\ \vdash \ p \lseg{l_1l_2} r   \ * \ r \listrep l_3
\end{align*}

%TODO talk about automation
%Let us sum up our comparison here. If we define predicates for partial data strucutres (like BST and linked list) using wand and quantifier, we will gain many properties for free.
%We could define them by recursion but then we need to prove those properties by induction.
%Also, such recursive definitions are very brittle and one must be very careful in all details or a useful rule becomes unsound.

\subsection{Disadvantages of using wand}

The proof theory of magic wand supports conjuncts-merging quite well. The derived rule \textsc{wandQ-elim} enables us to fill the hole of a partial data structure and get a complete one. The rule \textsc{wandQ-ver} enables us to shrink the holes of partial data structures. The rule \textsc{wandQ-hor} simply merges two holes into a larger one. The diagrams below illustrate these merging operations.
{\center
	\noindent\begin{tabular}{l}
~~~~~~~~~~~~~~~~		Fill the hole: \includegraphics[scale=0.65]{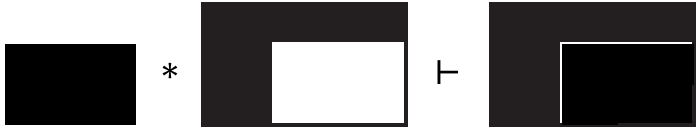}\\ [2ex]
~~~~~~~~~~~~~~~~		Vertical composition: \includegraphics[scale=0.65]{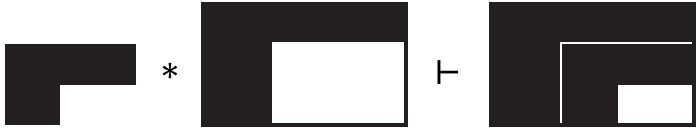}\\ [2ex]
~~~~~~~~~~~~~~~~		Horizontal composition: \includegraphics[scale=0.65]{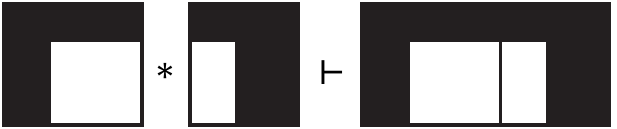}\\ [2ex]
	\end{tabular}
}

In contrast, the recursively defined $\mathrm{partial\_treebox\_rep^R}$ reveals more information about that partial tree but offers less support for merging. For example, we know that
\begin{align*}
& \mathrm{partial\_treebox\_rep^R}(L(P, x, v, t), r, i) \ \ \vdash \ \exists p. \ r \mapsto p \ * \ \fieldadd{p}{\mathsf{key}} \mapsto x \ * \ \top
\end{align*}
But we cannot prove any corresponding property about $\mathrm{partial\_treebox\_rep}$.
After all, \linebreak $\mathrm{partial\_treebox\_rep}$ is a weaker predicate---we can even prove: $$\mathrm{partial\_treebox\_rep^R}(P, r, i) \vdash \mathrm{partial\_treebox\_rep}(P, r, i)$$
But that magic wand expression precisely reveals the properties that we need in verification: the hole in it can be filled with another tree and it can also get merged with aother partial tree.

\subsection{The name: magic wand as frame}

In the execution of BST $\mathsf{insert}$'s loop body, the memory that the magic wand expression \linebreak $\mathrm{partial\_treebox\_rep}(P, p_0, p)$ describes is never touched by any C command.
Also, this expression is preserved as a separating conjunct in the assertions in the proof until getting merged with another conjunct in the end. Specifically, it gets merged with another $\mathrm{partial\_treebox\_rep}$ predicate by rule (\ref{eqn:ptr_ptr}) in two normal branches and it gets merged with a $\mathrm{treebox\_rep}$ by rule (\ref{eqn:tr_ptr}) in two \lstinline|return| branches.
In other words, although we do not explicitly use the frame rule in the proof, this magic wand expression acts as a frame.
Thus we call our proof a magic-wand-as-frame verification. This proof technique leverages the power of magic wand's proof theory for merging piece together.

%\section{Other proof variants}
%\input{variant}

\section{Related work and conclusion}\label{sec:concl}
\emph{Magic wand with local modification.}
Hobor and Villard \cite{hobor2013ramifications} use magic wand in their ramification theory in graph algorithm verification. Their proof rule \textsc{ramif} is used to connect local modification with its global effects.
\[
\inference[\textsc{ramif}]{\triple{L}{c}{L'} & G \vdash L * (L' \wand G') \\ \mathrm{FV}(L' \wand G') \cap \mathrm{ModV}(c) = \emptyset}{\triple{G}{c}{G'}}
\]
It can be treated as a special instance of magic-wand-as-frame proof. Using \textsc{wandQ-frame-intro}, \textsc{wandQ-frame-elim}, \textsc{frame} and Hoare logic's rule of consequence, we can prove a generalized version, \textsc{ramifQ}.
\[
\inference[\textsc{ramifQ}]{\triple{L}{c}{\exists x. L'(x)} & G \vdash L * \forall x. (L'(x) \wand G'(x)) \\ \mathrm{FV}(\forall x. (L'(x) \wand G'(x)) \cap \mathrm{ModV}(c) = \emptyset}{\triple{G}{c}{\exists x. G'(x)}}
\]

Moreover, Hobor and Villard provided the following rule for compositionality:
\[
\inference[]{G_1 \vdash L_1 * (L'_1 \wand G'_1) & G_2 \vdash L_2 * (L'_2 \wand G'_2) }{G_1 * G_2 \vdash (L_1 * L_2)* \left((L'_1 * L'_2) \wand (G'_1 * G'_2)\right)}
\]
This can be easily derived by \textsc{wandQ-frame-hor}.

\emph{Magic wand for partial data structure.}
Chargu\'eraud \cite{chargueraud2016higher} mentions in his paper that if the purpose of a partial tree is to fold back with the \emph{original} subtree (e.g. in BST look-up), magic wand can be used to describe that piece of memory.

Some shape analysis tools and theories \cite{Viper,DBLP:conf/tldi/MaedaSY11} can also support magic wand in verifying programs like BST-insert. They use wand assertions like $(\mathrm{tree\_shape}(p) \wand \mathrm{tree\_shape}(q) )$ to represent a partial tree.
Similar to Chargu\'eraud's claim, this local assertion is unchanged during the proof, although the local portion of memory may be modified.

Our method shows that \emph{even if the subtree is modified,} we can use a magic wand expression to describe a partial tree.
Moreover, using the quantifier, we can complete the verification even if we do not know the modification result at the beginning, e.g. if a new leaf's value is calculated through the path or passed from input.

\emph{Magic wand in interactive theorem proving.}
Iris has used magic wand heavily since Iris 3.0 \cite{krebbers2017essence}. They use magic wand and weakest-precondition (wp) to define their Hoare triple\footnote{This is only a simplified version for sequential programs.}:
$$\{P\}c\{Q\} \defeq \ \ \left( \vdash P \wand \mathrm{wp}(Q) \right)$$
They develop \emph{Iris Proof Mode} \cite{DBLP:conf/popl/KrebbersTB17} (which is a proof assistant inside another proof assistant) for proving such separation logic entailments in Coq, which simplifies the process of applying the adjoint property in the object language.

Magic-wand-as-frame limits the use of magic wand in a disciplined way. It is a light-weighted approach to make magic-wand-involved verification practical.

\emph{Separation logic for trees and lists:}
SmallFoot \cite{berdine05} verifies a shape analysis of a few linked list operations and tree operations.
Chargu\'eraud \cite{chargueraud2016higher} formalizes a series of separation logic verifications for high order linked lists and trees. They use recursively defined list segment and tree-with-a-hole instead of magic wand.  We have discussed their work in \S\ref{sec:comp}. 

\emph{Automatically verifying separation logic entailments for list segments.} If list-segment is recursively defined, rules like \ref{eqn:lseg_lseg} have to be proved by induction, which are hard to automate. Quang-Trung Ta \emph{et al.} \cite{DBLP:journals/pacmpl/TaLKC18} proposed a solver for building such induction proofs for list segments. But their algorithm is complicated. In comparison, if list-segments are defined as quantified wand expressions, corresponding proof rules can be proved by a simple tactic program.

In this paper, we demonstrate a Coq formalized verification of BST insert. Compared to the work of previous authors, our contributions are:
\begin{enumerate}
	\item We present a new proof technique: magic-wand-as-frame,
with its 5 rules (intro, elim, ver(tical composition),
hor(izontal composition), and refine).
\item We discover the power of magic wand in describing partial data structure. Our solution is the first one to handle both local modification (using quantifier) and loops that step through a data structure (by vertical combination). Its Coq formalization is also more concise because we can use Coq's high order logic and get partial tree's domain-specific theory for free.
\item We design a simple algorithm for proving partial data structure's elimination rules and vertical composition rules. Without magic-wand-as-frame, such rules needs to be proved by induction. An automated theorem prover for building induction proofs is complicated.
\item We formalize our proof in Coq and that formalization successfully uses those projects developed by other authors. Thanks to CompCert, Verifiable C and VFA, our Coq proof is actually an end-to-end correctness proof from top level specification to compiled assembly code.
\end{enumerate}

%We suggest that these uses in \emph{framing}, one can avoid the difficult proof
%theories and use our simple rules.

%\section{Review of separation logic}
%\input{preliminary}
%
%\section{Magic wand as frame} \label{sec:wand}
%\input{wand}
%
%\section{Q for quantifiers} \label{sec:wandQ}
%\input{wandQ}
%
%\section{P for program variables} \label{sec:wandP}
%\input{wandP}
%
%\section{Related work}
%\input{relate}
%
%\section{Conclusion}
%\input{concl}

\bibliographystyle{plain}
\bibliography{appel}

\begin{thebibliography}{10}

\bibitem{appel16:modsec}
Andrew~W. Appel.
\newblock Modular verification for computer security.
\newblock In {\em CSF 2016: 29th IEEE Computer Security Foundations Symposium},
  pages 1--8, June 2016.

\bibitem{appel17:vfa}
Andrew~W. Appel.
\newblock {\em Verified Functional Algorithms}, volume~3 of {\em Software
  Foundations}.
\newblock 2017.
\newblock URL: \url{softwarefoundations.org}.

\bibitem{appel14:plcc}
Andrew~W. Appel, Robert Dockins, Aquinas Hobor, Lennart Beringer, Josiah Dodds,
  Gordon Stewart, Sandrine Blazy, and Xavier Leroy.
\newblock {\em Program Logics for Certified Compilers}.
\newblock Cambridge, 2014.

\bibitem{berdine2004decidable}
Josh Berdine, Cristiano Calcagno, and Peter~W. O'Hearn.
\newblock A decidable fragment of separation logic.
\newblock In {\em International Conference on Foundations of Software
  Technology and Theoretical Computer Science}, pages 97--109. Springer, 2004.

\bibitem{berdine05}
Josh Berdine, Cristiano Calcagno, and Peter~W. O'{H}earn.
\newblock Smallfoot: Modular automatic assertion checking with separation
  logic.
\newblock In {\em Formal Methods for Components and Objects}, pages 115--135,
  2005.

\bibitem{blazy2009:clight}
Sandrine Blazy and Xavier Leroy.
\newblock Mechanized semantics for the clight subset of the c language.
\newblock {\em Journal of Automated Reasoning}, 43(3):263--288, Oct 2009.
\newblock URL: \url{https://doi.org/10.1007/s10817-009-9148-3}, \href
  {http://dx.doi.org/10.1007/s10817-009-9148-3}
  {\path{doi:10.1007/s10817-009-9148-3}}.

\bibitem{cao18:floyd}
Qinxiang Cao, Lennart Beringer, Samuel Gruetter, Josiah Dodds, and Andrew~W.
  Appel.
\newblock {VST-Floyd}: A separation logic tool to verify correctness of {C}
  programs.
\newblock {\em Journal of Automated Reasoning}, (to appear), 2018.

\bibitem{chargueraud2016higher}
Arthur Chargu{\'e}raud.
\newblock Higher-order representation predicates in separation logic.
\newblock In {\em Proceedings of the 5th ACM SIGPLAN Conference on Certified
  Programs and Proofs}, pages 3--14. ACM, 2016.

\bibitem{gu15:popl}
Ronghui Gu, J\'er\'emie Koenig, Tahina Ramananandro, Zhong Shao,
  Xiongnan~(Newman) Wu, Shu-Chun Weng, Haozhong Zhang, and Yu~Guo.
\newblock Deep specifications and certified abstraction layers.
\newblock In {\em 42nd ACM Symposium on Principles of Programming Languages
  (POPL'15)}, pages 595--608. ACM Press, January 2015.

\bibitem{hobor2013ramifications}
Aquinas Hobor and Jules Villard.
\newblock The ramifications of sharing in data structures.
\newblock In {\em POPL'13: Proceedings of the 40th Annual ACM SIGPLAN-SIGACT
  Symposium on Principles of Programming Languages}, pages 523--536. ACM,
  January 2013.

\bibitem{krebbers2017essence}
Robbert Krebbers, Ralf Jung, Ale{\v{s}} Bizjak, Jacques-Henri Jourdan, Derek
  Dreyer, and Lars Birkedal.
\newblock The essence of higher-order concurrent separation logic.
\newblock In {\em European Symposium on Programming}, pages 696--723. Springer,
  2017.

\bibitem{DBLP:conf/popl/KrebbersTB17}
Robbert Krebbers, Amin Timany, and Lars Birkedal.
\newblock Interactive proofs in higher-order concurrent separation logic.
\newblock In Giuseppe Castagna and Andrew~D. Gordon, editors, {\em Proceedings
  of the 44th {ACM} {SIGPLAN} Symposium on Principles of Programming Languages,
  {POPL} 2017, Paris, France, January 18-20, 2017}, pages 205--217. {ACM},
  2017.
\newblock URL: \url{http://dl.acm.org/citation.cfm?id=3009855}, \href
  {http://dx.doi.org/10.1145/3009837} {\path{doi:10.1145/3009837}}.

\bibitem{DBLP:conf/tldi/MaedaSY11}
Toshiyuki Maeda, Haruki Sato, and Akinori Yonezawa.
\newblock Extended alias type system using separating implication.
\newblock In Stephanie Weirich and Derek Dreyer, editors, {\em Proceedings of
  {TLDI} 2011: 2011 {ACM} {SIGPLAN} International Workshop on Types in
  Languages Design and Implementation, Austin, TX, USA, January 25, 2011},
  pages 29--42. {ACM}, 2011.
\newblock URL: \url{http://doi.acm.org/10.1145/1929553.1929559}, \href
  {http://dx.doi.org/10.1145/1929553.1929559}
  {\path{doi:10.1145/1929553.1929559}}.

\bibitem{Viper}
Peter M{\"{u}}ller, Malte Schwerhoff, and Alexander~J. Summers.
\newblock Viper: {A} {Verification Infrastructure for Permission-Based
  Reasoning}.
\newblock In Barbara Jobstmann and K.~Rustan~M. Leino, editors, {\em
  Verification, Model Checking, and Abstract Interpretation - 17th
  International Conference, {VMCAI} 2016, St. Petersburg, FL, USA, January
  17-19, 2016. Proceedings}, volume 9583 of {\em Lecture Notes in Computer
  Science}, pages 41--62. Springer, 2016.
\newblock URL: \url{https://doi.org/10.1007/978-3-662-49122-5\_2}, \href
  {http://dx.doi.org/10.1007/978-3-662-49122-5\_2}
  {\path{doi:10.1007/978-3-662-49122-5\_2}}.

\bibitem{reynolds02}
John Reynolds.
\newblock Separation logic: A logic for shared mutable data structures.
\newblock In {\em LICS 2002: IEEE Symposium on Logic in Computer Science},
  pages 55--74, July 2002.

\bibitem{rowling97}
J.~K. Rowling.
\newblock {\em Harry Potter and the Philosopher's Stone}.
\newblock Bloomsbury Childrens, London, 1997.

\bibitem{DBLP:journals/pacmpl/TaLKC18}
Quang{-}Trung Ta, Ton~Chanh Le, Siau{-}Cheng Khoo, and Wei{-}Ngan Chin.
\newblock Automated lemma synthesis in symbolic-heap separation logic.
\newblock {\em {PACMPL}}, 2({POPL}):9:1--9:29, 2018.
\newblock URL: \url{https://doi.org/10.1145/3158097}, \href
  {http://dx.doi.org/10.1145/3158097} {\path{doi:10.1145/3158097}}.

\end{thebibliography}

\end{document}